\documentclass[aps,prl,preprint,showpacs,amsmath,amssymb,floatfix]{revtex4}
\usepackage{graphicx}
\usepackage{times}
\begin{document}

\preprint{Submitted to Physical Review Letters}

\title{Equilibrium Simulation of the Slip Coefficient in Nanoscale Pores}

\author{Vlad P. Sokhan}
\email{vlad.sokhan@npl.co.uk}
\affiliation{National Physical Laboratory, Hampton Rd,
             Teddington TW11 0LW, United Kingdom}

\author{Nicholas Quirke}
\altaffiliation[Also at ]{Department of Physics, University College 
                          Dublin, Ireland}
\affiliation{Department of Chemistry, Imperial College London, London NW7 2AY,
             United Kingdom}

\date{\today}

\begin{abstract}
Accurate prediction of interfacial slip in nanoscale channels is required by 
many microfluidic applications. Existing hydrodynamic solutions based on 
Maxwellian boundary conditions include an empirical parameter that depends on 
material properties and pore dimensions.
This paper presents a derivation of a new expression for the slip coefficient 
that is not based on the assumptions concerning the details of solid-fluid 
collisions and whose parameters are obtainable from \textit{equilibrium} simulation. 
The results for the slip coefficient and flow rates are in good agreement 
with non-equilibrium molecular dynamics simulation.
\end{abstract}

\pacs{PACS numbers: 47.15.gm, 47.11.Mn, 05.20.Jj, 47.90.+a, 66.20.+d, 68.08.--p \\
{\copyright~~Crown copyright, 2007} }

\maketitle

Many recent experiments on nanoscale fluid flow of Newtonian liquids
\cite{el07,ne05} and related simulation studies
\cite{bb99,zhu01,vps01,cb03} support Maxwell's prediction of the
possibility of molecular slip at a gas-solid interface \cite{max79}.
It is known to exist where the Knudsen number, defined as the dimensionless
ratio of molecular free path to some characteristic length,
$Kn=\lambda /L,$ is non-negligible, and is usually associated with 
low densities where $\lambda$ is large. However, in narrow capillaries
(where $L$ is small), slip can be observed even at liquid densities.
In the general case it is characterized by a slip coefficient, $l_{\text{s}}$ 
\cite{max79,ce69}, which in the absence of temperature gradients
relates the collective molecular velocity at the wall, the slip velocity, 
to the shear rate, ${\bf u}_{\text{s}}=l_{\text{s}}\nabla{\bf u}$ \cite{max79}. 
Using kinetic theory Maxwell derived a microscopic expression for the slip 
coefficient \cite{max79}, that can be written as
\begin{equation}
 \label{eq1}
  l_{\text{s}}=\lambda \left( {\frac{2}{\alpha}-1} \right),
\end{equation}
where $\lambda={2\eta}/{\rho\bar{c}}$ is the mean free path, $\eta$ is the 
shear viscosity, $\rho$ is mass density of the fluid, and $\bar{c}$ is the 
mean speed of the molecules. Considering only two types of wall collision, 
specular and diffuse (Knudsen) reflection, he introduced a coefficient
$\alpha$ that defines a fraction of specularly reflected molecules. In more broad 
sense, $\alpha$ defines the fraction of the flux of tangential momentum transmitted 
in collisions and is often called the `accommodation coefficient' (TMAC) \cite{ce69}.
Its value is defined by the details of the solid-fluid interactions and (\ref{eq1}) 
implies finite slip even for purely diffusive reflections ($\alpha =1$). 
For specularly reflecting surface the slip coefficient diverges since the fluid 
cannot grip onto the surface.

\begin{figure} 
\includegraphics*[width=0.9\linewidth]{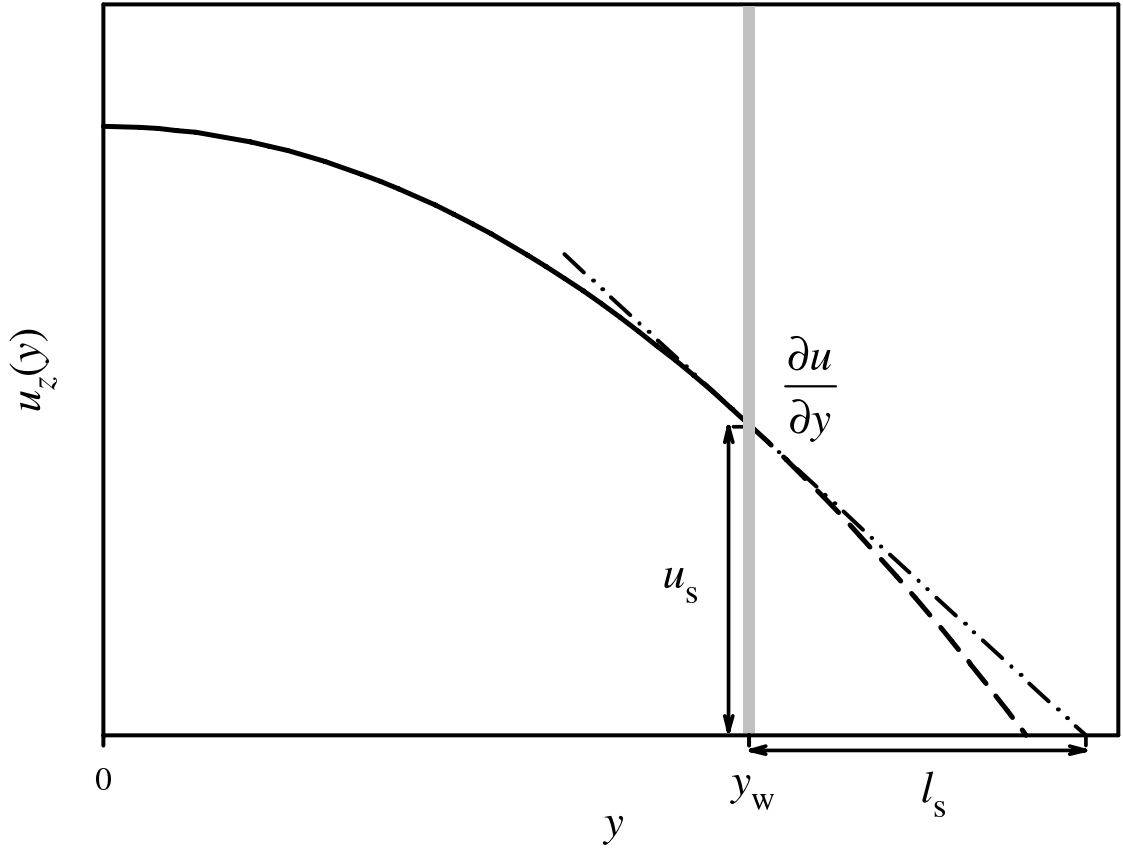}
\caption{\label{fig01}Full line, velocity profile for the plane Poiseuille flow with 
surface slip; dashed line, extrapolated velocity profile. Vertical gray line 
marks the position of the solid. Dash-dotted line denotes the velocity 
gradient at the wall. Also shown are slip velocity, $u_{\text{s}},$ and slip 
coefficient, $l_{\text{s}}.$}
\end{figure}

Slip in nanoscale fluid flow depend on many parameters inculding surface roughness \cite{cb03}, 
electric properties of the interface \cite{jo06}, wetting conditions \cite{vps01}, 
chemical patterning of the surface \cite{qi05}, and is a nonlinear function 
of the dynamic state \cite{tt97}.
Application of the generalized Navier-Stokes (NS) hydrodynamics to problems 
of fluid flow on the nanoscale is an attractive but highly nontrivial 
task even in simple cases such as plane Poiseuille flow. An accurate 
solution requires a knowledge of the material parameters of the fluid as a 
function of local density which deviates from its bulk value in the proximity 
of the interface. As a result, the velocity profile in narrow pores 
also deviates from the macroscopic prediction \cite{tg00}.
The usual approach is to regard the fluid as incompressible and to replace 
the complex non-uniform flow problem plus simple no-slip boundary conditions 
with a simple flow problem but with boundary conditions that has been called 
`exceedingly difficult' for theoretical investigation \cite{ce69}, p.~96. 
The discontinuity in the flow field is introduced in this approach in the 
same way the surface excess was introduced by Gibbs in his treatment of 
interface boundaries at equilibrium as illustrated in Fig.~\ref{fig01} for the 
case of plane Poiseuille flow in $z$ direction 
(${\rm{\bf u}}=\left\{{0,0,u_z}\right\})$ between 
walls at $y=\pm h$, where the symmetry of the solution was taken into 
account and only one half was plotted. We emphasize that the finite slip 
$u_{\text{s}}$ that appears in this approach is a purely artificial device 
introduced only to match the approximate solution with the solution of full 
problem in the middle part of the channel. The full solution for the 
continuum velocity field does decay to zero at the wall, as required by 
continuity of stresses in classical hydrodynamics \cite{ba70}. 
It should be noted however that the limiting value of the velocity field 
cannot be observed in any experiment or in a statistical particle-based 
simulation with a continuous solid-fluid potential since the velocity field, 
which is defined everywhere the fluid density is non-zero, cannot be 
measured from particle velocities in regions where particles are not 
observed due to finite sampling of the statistical ensemble, \textit{i.\,e.}
where the particle Boltzmann factor is diminishingly small but non-zero.

There were a number of attempts to estimate the TMAC 
from kinetic theory \cite{nb06,am06} and using molecular dynamics simulation 
of simple liquids \cite{vps01,vps02,ar03}. However, the results obtained by different 
groups and using different methods are inconcluisive \cite{el07}, with large 
scatter in obtained values of slip coefficient spanning two orders of magnitude.
The agreement between recent experiments with various surfaces \cite{lm99,ar01,sa01,cao05}, 
and and theory is also poor, with experimental values for $\alpha$ generally higher 
than the corresponding theoretical expectations \cite{lm99,vps02,am06}.
One of the limitations of the original Maxwell model is that it was developed to solve 
the half-space flow, or Kramers, problem \cite{ce69} in the dilute gas regime and faces 
problems describing flow in pores of finite width. It is assumed that the TMAC is 
a local property independent of the pore width. 
The following argument however demonstrates why it could not be so. 
Viscous wall stress in plane Poiseuille flow that is proportional to the velocity 
gradient at the wall scales as the pore width for the flow of gas of the same density.
This stress is transmitted to the wall through collisions that are independent of the
pore width and therefore the momentum transferred to the wall in each collision and
consequently the TMAC depends on the pore width.

In this paper we present a method of calculating the slip coefficient from 
equilibrium simulation that does not require assumption about the type of wall 
collisions avoiding thus the necessity of calculating the TMAC. We illustrate 
the method on a simple case of gravity driven plane Poiseuille flow in a pore 
of width $H=2h$ and with acceleration due to an external field 
${\rm{\bf g}}=\left\{{0,0,g}\right\}$ as sketched in Fig.~\ref{fig01}. 
The method can be extended to a more general case of Poiseuille 
flow induced by a pressure gradient using the equivalence between the 
pressure gradient and gravity-driven force in the direction of flow 
\cite{ba70}. It is convenient to reformulate the slip velocity 
problem (the Dirichlet boundary condition) in terms of interfacial viscosity 
(Neumann boundary condition), and the lateral wall stress. This idea 
goes back to Navier \cite{na27} who obtained the boundary condition 
for the velocity field on the basis of particle arguments (cf.\ last eqn.\ on 
p.~415 in Ref.~\cite{na27}) as,
\begin{equation}
 \label{eq2}
  \eta \left. {\frac{\partial u_z (y)}{\partial y}} \right|_{y_w } =\beta 
  u_z \left( y_w \right)
\end{equation}
where his parameter $\beta$ is related to the interfacial shear viscosity $\eta$
via $l_{\text{s}}=\eta/\beta.$ Note that for non-linear velocity profiles,
as for Poiseuille flow, the slip coefficient, $l_{\text{s}},$
(shown in figure) is different from the slip length, defined as the 
distance from the wall where the extrapolated velocity profile vanishes. 
However, if the difference between them is small, simple geometric 
consideration allows us to establish the relationship between the two types 
of boundary condition.

Neglecting viscous dissipation, the wall shear stress in terms of the external 
driving force acting on fluid particles, which in this case is simply 
$\sigma_{yz} = \rho gh$ \cite{vps05}, can be equated with the Stokesian drag
force per unit area exerted on the wall by the moving fluid \cite{vps04},
\begin{equation}
 \label{eq3}
  \sigma_{yz}=-\frac{F_S}{A}=\frac{Mu}{2A\tau}\equiv \frac{\rho uh}{\tau},
\end{equation}
where $M$ is the total mass of the fluid in the pore; $u\equiv h^{-1}\int_0^h 
{u_z(y)dy}$ is the mean fluid velocity, and the relaxation time $\tau$ can be 
calculated in molecular dynamics simulation from the Langevin equation for 
the fluid subsystem considered as a single Brownian particle using 
fluctuation-dissipation theorem. It has been shown recently \cite{vps04}
that the fluid velocity autocorrelation function decays exponentially,
\begin{equation}
 \label{eq4}
  C(t)=M^{-1}kT\exp \left(-t/\tau\right).
\end{equation}
This provides a simple way to determine $\tau$ in an equilibrium simulation
by fitting a one-parameter exponential to velocity autocorrelation data. From 
the two expressions for the wall shear stress, a simple relationship between 
the fluid velocity and the acceleration due to the external force,
\begin{equation}
 \label{eq5}
  u=\tau g.
\end{equation}
can be established. This surprisingly simple result shows that within the 
limits of linear regime \cite{tt97} the rate of fluid flow in 
non-equilibrium steady state can be estimated from the characteristic time 
of the decay of fluctuations at equilibrium. Using it, we can also estimate 
both the slip velocity and slip coefficient. Since the hydrodynamic solution 
in this case is given by a quadratic velocity profile with slip,
\begin{equation}
 \label{eq6}
  u_z (y)=\frac{\rho g}{2\eta}\left( {h^2-y^2} \right)+u_{\text{s}},
\end{equation}
using (\ref{eq5}) and the definition of $u$, one obtains for slip velocity
\begin{equation}
 \label{eq7}
  u_{\text{s}} =\left( {\tau -\frac{\rho h^2}{3\eta }} \right)g,
\end{equation}
and by substituting it into the definition of the slip coefficient, 
$l_{\text{s}}=-\left. {{\partial u}/{\partial y}}\right|_h u_{\text{s}},$
one obtains finally
\begin{equation}
 \label{eq8}
  l_{\text{s}} =\frac{\tau\eta}{\rho h}-\frac{h}{3}.
\end{equation}
This is the main result. It shows that the slip coefficient is independent of the
external force (flux), but nonlinearly depends on the pore width, both directly
and indirectly through the relaxation time $\tau\equiv\tau(h)$.
The connection with the Maxwell's result can be established by using the 
relationship between the TMAC and relaxation time, $f_0,$ $\tau=(f_0\alpha)^{-1}$ 
\cite{vps04}. Taking the kinetic theory expressions for the wall collision 
frequency per particle, $f_0=\bar{c}/4h$ and viscosity, obtain the expression for 
the slip coefficient
\begin{equation}
 \label{eq9}
  l_{\text{s}} =\lambda\frac{2}{\alpha}-\frac{h}{3}
\end{equation}
that can be directly compared with Maxwell's result (\ref{eq1}). We note here that 
usage of kinetic theory expression for $f_0$ introduces a significant error at 
liquid densities, and it was used in deriving (\ref{eq9}) only to make the 
same level of approximation as was in deriving (\ref{eq1}).

In order to test the accuracy of (\ref{eq8}), two series of equilibrium molecular 
dynamics runs were performed. There are two ways of controlling the Knudsen number
in simulation: either by changing the mean density of the system (the mean free path) 
or the pore width (the characteristic length), and we used both.
In the first series, the relaxation time $\tau$ was calculated 
as a function of fluid density for a pore of fixed width. In the second, the pore width 
was varied while the normal wall pressure was kept constant.
The slip coefficient estimated using (\ref{eq8}) was compared with the values obtained
directly in the parallel set of nonequilibrium molecular dynamics (NEMD) simulations of 
the Poiseuille flow for the same systems.
The system consists of supercritical Lennard-Jones (LJ) fluid, confined 
between two walls modeled by a rigid
triangular lattice of atoms situated at $y_{\text{w}} =\pm h,$ and periodically 
replicated along $x$ and $z$ axes to avoid edge effects. All interactions
in the system were of the LJ form, $U(r)=4\varepsilon\left[{\left(\sigma/r
\right)^{12}-\left(\sigma/r\right)^6}\right],$ where $\varepsilon$ and $\sigma$
are the usual energy and length parameters. Their values for the fluid-fluid
interactions define the corresponding scales, and in the following reduced
units \cite{at87} are used, denoted by the astersk. The solid-fluid interaction parameters in 
these units were taken as $\varepsilon_{\text{sf}}=0.4348\varepsilon$ and 
$\sigma_{\text{sf}}=0.9462\sigma,$ which is appropriate for methane gas 
between carbon `rare-gas walls'. The surface number density of the solid was 
$n_{\text{s}}\sigma^2=1.105$ as in our earlier study \cite{vps04} 
where more details about the system and the numerical scheme can be found. 
All molecular dynamics calculations were performed using the classical 
molecular dynamics software package MDL \cite{mdl}. 
The nonequilibrium steady-state conditions were realized by placing the fluid
in a uniform external field parallel to the walls and coupling all fluid 
degrees of freedom to a Nos\'{e}-Hoover thermostat at $T=2.026\,\varepsilon 
k^{-1},$ where $k$ is the Boltzmann constant. The simulation cell was of 
dimensions $20.181\sigma \times H\times L_z,$ where the dimension in the 
flow direction, $L_z,$ was scaled with the density to keep the number of 
particles in the system around 2500. The timestep was $\Delta t^{\ast} 
=7.28\times 10^{-3}$ in LJ units and the integration time in each case was 
not less than $3.64\times 10^5$ (50M steps).
The value of the acceleration due to the external force was varied between 
$g^{\ast} =4\cdot 10^{-3}$ and $g^{\ast} =0.04$ in order to keep the fluid 
velocity below $u^{\ast}=0.2.$ The steady states reported here are known to 
be well within the Newtonian regime \cite{bb00} and the slip velocity in 
the linear regime \cite{tt97}. To simplify the comparison we scaled 
all flow related properties to a common value $g^{\ast} =4.95\cdot 10^{-3}$. 

In the first series three pore widths were considered, $H=5.328\sigma,$ 
$10.499\sigma,$ and $20.997\sigma;$ (in the future denoted as $5\sigma$, 
$10\sigma$, and $20\sigma$ for brevity) and several number densities 
ranging from $n^{\ast}=0.02$ ($n^{\ast}=m^{-1}\rho\sigma^3)$, corresponding to 
the rarefied fluid at $Kn=2.34$ for a narrow pore (where we used the kinetic 
theory expression for the mean free path and the hard sphere diameter for LJ 
\cite{he88}), to a dense state of $n^{\ast}=0.8$ that for a wide pore 
gives $Kn=0.015$, thus spanning more than two orders of magnitude of Knudsen 
numbers. In all cases the Reynolds number defined as the ratio of 
inertial and viscous forces, $Re=\rho uH/\eta,$ was kept small, $Re<10$.

\begin{figure} 
\includegraphics*[width=0.9\linewidth]{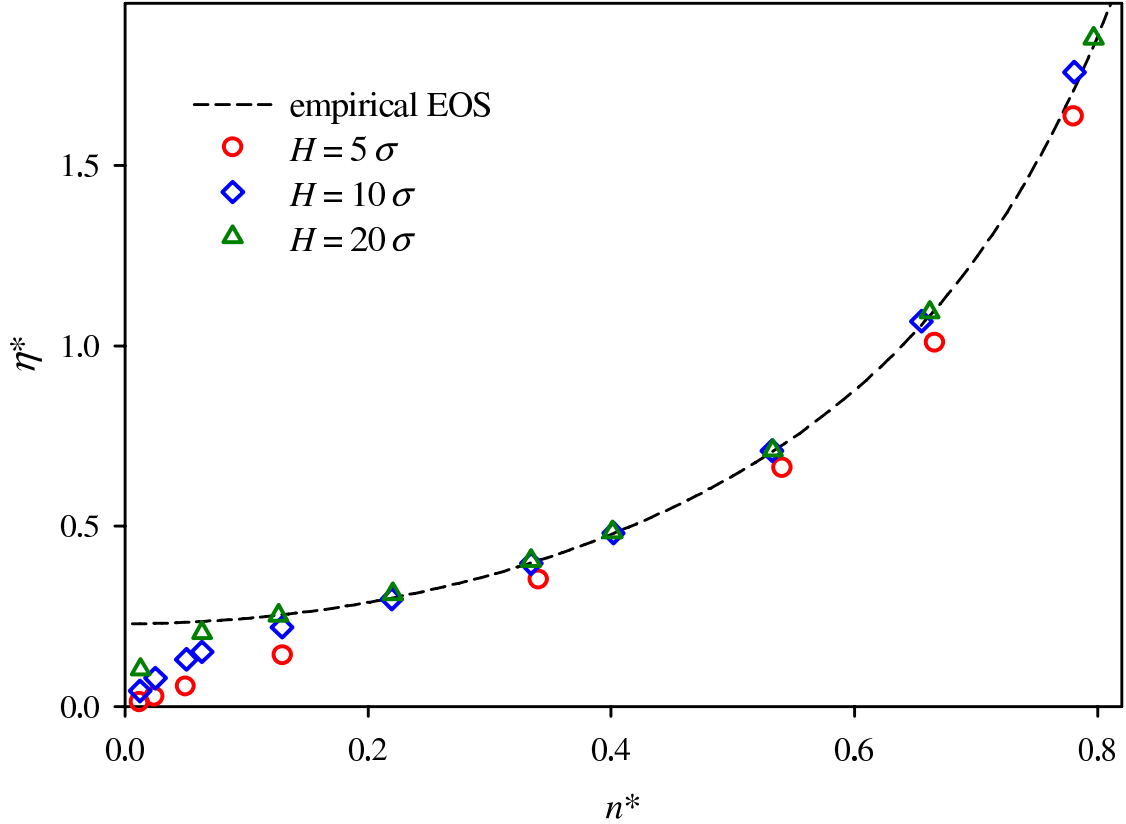}
\caption{\label{fig02}(color online). Comparison of the calculated shear viscosities 
                for three pore widths as a function of reduced number density in the 
                centre of pore, symbols, with empirical EOS for bulk fluid, dashed line.}
\end{figure}

In order to calculate slip velocity using (\ref{eq8}) we need the estimates 
of the shear viscosity in the channel. It can be calculated in equilibrium
moleuclar dynamics \textit{e.g.} from the stress autocorrelation function
using Green--Kubo relations \cite{ml04}. For the bulk LJ fluid it has been 
recently accurately estimated in molecular dynamics simulation \cite{ml04}. 
To simplify the calculations we used an empirical equation of state for the 
viscosity \cite{lv05} at a density equal to the mean density in the central 
part of the channel, where it is uniform to fit the simulated data of \cite{ml04} 
at the required temperature. The accuracy of this procedure was estimated by
comparing the obtained values with values calculated from the NS hydrodynamics 
using the quadratic fit to NEMD velocity profiles. 
The results for three pore widths are presented in Fig.~\ref{fig02}. 
The agreement for the two wider pores is excelent. Deviation from the bulk 
values for the third pore, $H=5\sigma$, is due to the overlap of adsorbed 
layers at two surfaces and as a consequence, to inaccuracy in determination of
the corresponding bulk density. At low densities the viscosity markedly deviates 
from bulk values for all three pores when $Kn\ge 1$.

\begin{figure} 
\includegraphics*[width=\linewidth]{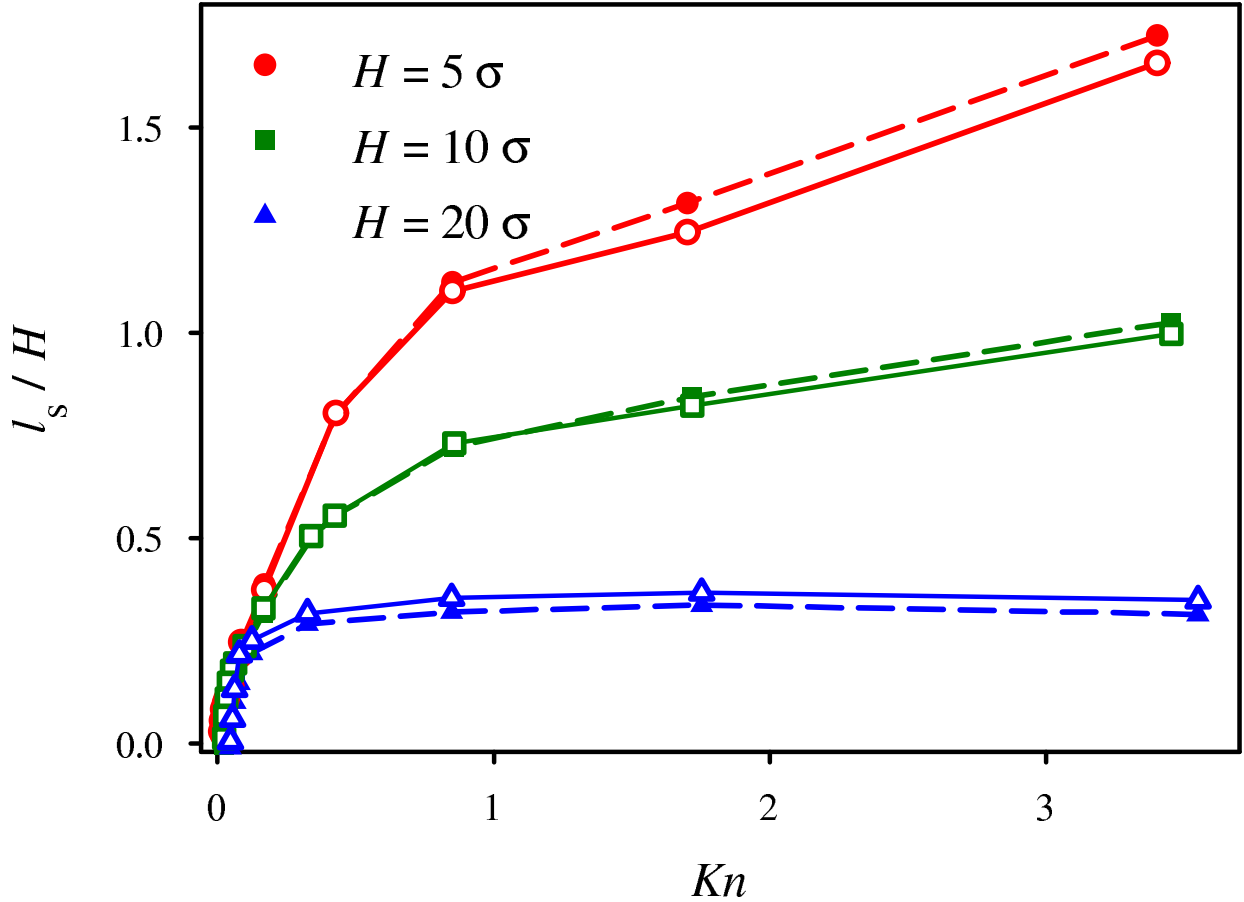}
\caption{\label{fig03}(color online). Slip coefficients calculated directly in NEMD 
                (open symbols and full lines) and estimated from the relaxation 
                time (eqn.\ (\ref{eq8})), filled symbols and dashed lines) as a function 
                of Knudsen number for three pore widths. Lines are drawn to guide the eye.}
\end{figure}

The ratio of the slip coefficients to the pore width, calculated in NEMD and estimated from 
the relaxation times and bulk viscosities at densities equal to that in the middle of pore 
using (\ref{eq8}) are compared on Fig.~\ref{fig03} for three pore widths.
The relaxation time was estimated from the exponential fit to the collective velocity
autocorrelation function calculated in equilibrium molecular dynamics simulaton using the 
procedure described in \cite{vps04}. The statistical uncertainty in all cases is of the order 
of symbol size and is slightly higher for wider pores since the longer relaxation times in 
this case require more accurate estimation of the velocity autocorrelation time at long 
correlation times. For all pores the slip coeficients calculated using the two routes agree 
within statistical uncertainties. 
For a given pore, both (\ref{eq1}) and (\ref{eq9}) predict linear dependence of the slip length on
the Knudsen number with zero and negative offset, correspondingly. For two wider pores the behaviour
is observed at high Knudsen numbers, $Kn>1,$ but with positive offset that increases with pore width.
Surprisingly, for the narrowest pore, the slip length appears to be independent of the Knudsen number 
(density) for $Kn>1.$ Since density enters only the first term in (\ref{eq8}), this reslt indicate that
in the low density regime the relxation time in narrow pores increases linearly with density.

\begin{figure} 
\includegraphics*[width=\linewidth]{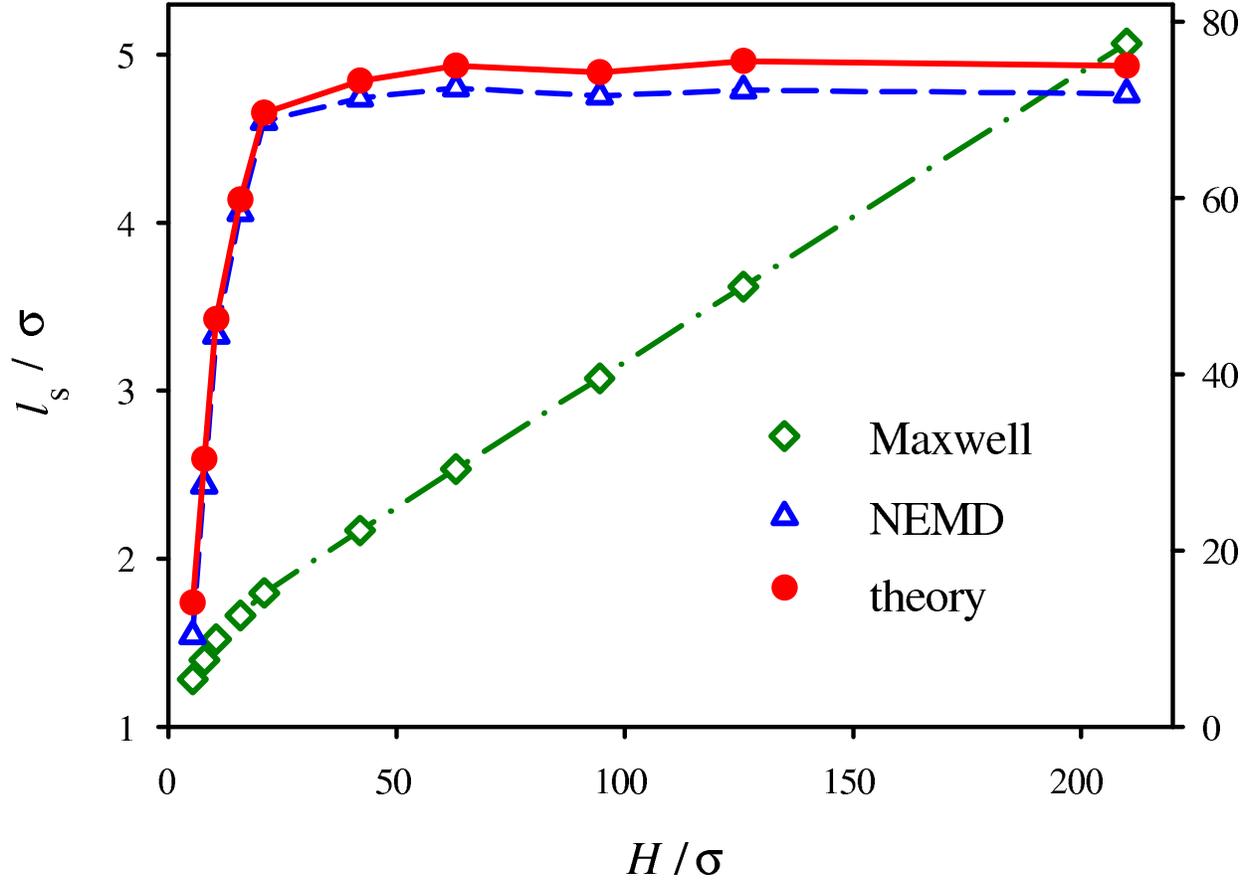}
\caption{\label{fig04}(color online). Slip coefficients (left scale) calculated directly in NEMD, 
                triangles, and estimated from the relaxation time (eqn. (\ref{eq8})), 
                circles, as a function of pore width. For comparison, slip coefficients
                estimated from the Maxwell's theory, eqn. (\ref{eq1}), diamonds,
                are also shown (right scale). Lines are drawn to guide the eye.}
\end{figure}

The results also show that for the systems with the same Knudsen number the slip length increases 
with the pore width. In order to establish whether there is a limiting value of the slip coefficient
a second series of calculations was performed at the normal pressure that corresponds to reduced
density of $n^{\ast}=0.125$ up to the pore width $H=210\sigma.$ Obtained values of the reduced 
slip coefficient are compared on Fig.~\ref{fig04} with the Maxwell's results using (\ref{eq1}).
The results estimated using (\ref{eq9}) agree with those obtained directly in NEMD within statistical 
uncertainty. They show that at about $H=25\sigma$ the slip coefficient reaches the limiting value
$l_{\text{s}}=4.77(2)\sigma$. Maxwell's theory, on the contrary, predicts linear scaling of the slip 
length with pore width and for the widest pore studied it gives the value $l_{\text{s}}=77.5\sigma$
(note the difference in scales).

\begin{acknowledgments}
This work was funded as part of the National Physical Laboratory's Strategic 
Research Programme and EPSRC under grant GR/N64809/01.
\end{acknowledgments}

\end{document}